# Direct Measurement of Penetration Length in Ultra-Thin and/or Mesoscopic Superconducting Structures

L. Hao

*National Physical Laboratory, Teddington, TW11 0LW, United Kingdom;*

J.C. Macfarlane

*Department of Physics, University of Strathclyde, Glasgow G4 0NG, United Kingdom*

J.C. Gallop

*National Physical Laboratory, Teddington, TW11 0LW, United Kingdom;*

and S.K.H. Lam

*CSIRO, Lindfield 2070, Australia.*

As the dimensions of thin superconducting structures become comparable with or less than the penetration length of magnetic flux into the structures, it becomes increasingly necessary to devise experimental tests of available theoretical models. One approach which we shall describe, enables penetration lengths to be derived from the measurement of the effective area of planar, thin-film structures with linear dimensions in the range 1 to 100μm. The effective area is defined by measurement of the inductive coupling of the structures to dc or low-frequency magnetic fields. The structures described consist of two parts: (1) An ultra-thin annular superconducting film with transition temperature $T_{ca}$ ("washer"); and (2) surrounding the washer is a superconducting ring with transition temperature $T_{cs}$. Because the films are prepared in such a way that $T_{ca} < T_{cs}$, the ring-washer combination acts as a dc SQUID (Superconducting Quantum Interference Device) up to and beyond $T_{ca}$, enabling the effective area of the washer to be measured over a wide temperature range. Results for the temperature dependence of the Pearl penetration length $\Lambda(T)$, derived directly from measurements of the effective area, are compared both with theory and with other experimental data. Whereas alternative methods may be restricted to narrow-band, high-frequency fields and require sample dimensions of order 10 mm or greater, the method is inherently broad-band and is applicable to dimensions $\geq 1$ μm.

# I. INTRODUCTION

The magnetic properties of superconducting planar structures such as thin-film disks or rings in magnetic fields perpendicular to their plane are of interest, both from theoretical and practical considerations. Recently, calculations by Brandt et al. [1] for the inductance and effective area of thin-film disks and rings have been extended from the ideally-screening Meissner state to include the more practically-interesting cases where the London penetration depth $\lambda$ or, equivalently, the two-dimensional (Pearl) screening length $\Lambda = \lambda^2/d$, is comparable to or greater than the film thickness $d$. In the case of the ideally-screening superconducting ring, magnetic flux is entirely excluded, whereas when $\lambda \geq d$ magnetic flux is able to penetrate into the aperture of the ring, as well as into the film itself. In many practical situations, where the film thickness may be of the order 10 - 20 nm, this condition prevails at temperatures between $T_c/2$ and $T_c$, where $T_c$ is the superconducting transition temperature. In published work, measurements of the penetration lengths in thin superconducting films have been based on the 2-coil method of Fiory et al.[2] or variations thereof (eg Claassen et al[3]., Wang et al.[4], ) where the lateral dimensions of the film are necessarily on the order of 10 mm due to the dimensions of practical wirewound coils. Other techniques based on measuring the change in resonant frequency of a microwave cavity, and on measuring the transmission of microwaves through the film, have also been recently described (Gubin et al., [5]). Again, these methods require sample dimensions of at least several millimetres, and are necessarily constrained to working in the high-frequency range. With a growing need to reduce the dimensions of superconducting devices into the sub-micrometre regime (e.g. for quantum computing or single particle counting applications) it is necessary to consider new, non-invasive, in-situ techniques for the measurement of the temperature-dependent magnetic properties of ultra-thin superconducting structures. The technique reported here has been developed with such an application in mind, which will be described in Section VI. In contrast to the works referred to above, the methodology we have developed is suited to measuring the penetration length in ultra-thin-film structures of micrometer dimensions, and is inherently broad-band (dc to MHz) in its frequency response. The particular geometry chosen for this study, in recognition of its wide use in the design of SQUIDs (Superconducting Quantum Interference Devices), is that of a circular or polygonal lamina with a central hole and a radial slit, conventionally referred to as a slit washer. Other geometries are also amenable to the method, as shown by ongoing studies which we shall report in the future.

## II. THEORETICAL BACKGROUND

First we note that conventional "rule-of-thumb" expressions (e.g Ketchen, [6]), for the inductance and effective area of washer-type SQUIDs, do not take account of flux penetration into the thin-film structures. These approximations will accordingly give unreliable estimates of the effective area and inductance in cases where the film thickness $d \leq \lambda < \Lambda$. The Pearl length $\Lambda$ is related to $\lambda$ and is given by

$$\Lambda = (\lambda_0^2/d)/(1-(T/T_c)^4). \qquad (1)$$

where $\lambda_0$ is the penetration depth at $T$=0K, and we have assumed a London model for the temperature dependence of the penetration depth $\lambda(T)$.

Although Ketchen [6] described numerical solutions for particular cases, more detailed calculations and graphs of inductance and effective area as a function of penetration depth in this regime have been recently published [1]. Experimental data for the penetration depth at GHz frequencies have also been reported [5]. We compare our experimental results with data from both papers [1,5] and show that our method is applicable to much smaller device sizes.

In practice, the experimental quantity that is directly measured in quantum interference devices (such as SQUIDs) is the effective area, given by the increment $\Delta B$ in the applied perpendicular magnetic field corresponding to one period in the SQUID's voltage-field characteristic:

$$A_{eff} = \Phi_0/\Delta B \qquad (2)$$

where $\Phi_0 = h/2e$ is the magnetic flux quantum.

## III. EXPERIMENTAL APPROACH

A direct measurement of the temperature dependence of a conventional SQUID's effective area or self-inductance is hindered by the fact that the critical currents of the Josephson junctions which determine the SQUID performance are themselves strongly temperature- and field- dependent, so that variations in their transport properties will alter or obscure the inductance variations. To avoid this problem, we have introduced a modified type of device (e.g. fig. 1a) where an outer superconducting ring with transition temperature $T_{cs}$, linewidth $s$ and outer diameter $2b$ forms a dc SQUID. Within the ring, a slit-washer structure with inner hole diameter $2a$ is deposited, consisting of a superconducting film of thickness $d$ having a transition temperature $T_{ca} < T_{cs}$.

On the assumption that for $T < T_{ca}$, $A_{eff}$ and penetration length $\Lambda$ are related by the Ketchen formula [6] (modified to include a finite penetration length $\Lambda(T)$:

$$A_{eff} = \pi(a + \Lambda(T))b \qquad (3)$$

the temperature dependence of $\Lambda$ can be derived as

$$\Lambda(T) = (A_{eff} / \pi b) - a. \qquad (4)$$

At the other extreme, where $T_{ca} < T < T_{cs}$ the effective area will approach the geometrical area

$$A_g = \pi(b-s)^2 \qquad (5)$$

An example of the device is shown in the scanning-electron micrograph Fig.1(b). In this example, the outer superconducting ring (octagonal in this example, but assumed to behave similarly to a circular geometry) is made from Nb of thickness $d \sim$ 20-100 nm, diameter $2b$ = 50 μm, linewidth $s$ = 5 μm and incorporates two weak links of approximate dimensions 80 x 100 nm. Within the ring, the slit-washer structure has inner hole diameter $2a$ and consists of a superconducting Nb film of thickness $d \sim$ 15 nm having a transition temperature $T_{ca} < T_{cs}$.

We would therefore expect the SQUID-washer combination to resemble a conventional SQUID of the Ketchen [6] type at temperatures $T < 0.5 T_{ca} \ll T_{cs}$, while at temperatures $0.5 T_{ca} < T < T_{cs}$, the effective area will increase with temperature, up to some limit at which the washer (but not the SQUID ring) is entirely in a non-superconducting state, i.e. when $T_{ca}<T<T_{cs}$ , only the perimeter is superconducting, and the area will approach the geometric area, $A_g$~1200 µm$^2$ (Eq. 5). In the lower limit, at $T<<T_{ca}$ when the entire structure is in the ideal Meissner state, Eq. (3) (which also accounts for the effects of flux focussing) gives an estimated effective area of ~350 µm$^2$. The slot provides an additional area of ~100 µm$^2$, making a total area of ~450 µm$^2$. The structures are prepared from sputtered Nb films on oxidized silicon wafers. The SQUID rings and absorber patch are prepared by conventional lithography, whereas the nano-bridge junctions are made either by focussed-ion-beam [7] or electron-beam techniques [8]. This approach enables a valuable simplification of the fabrication process, in that the nano-bridges can be introduced at a late stage directly into the previously-deposited single-layer Nb SQUID ring.

We chose not to use conventional SQUIDs which incorporate Josephson tunnel junctions (of order 4µm$^2$ in area) because, as discussed above, their sensitivity to applied fields would complicate the interpretation of the effective area data. The nano-bridge weak links, on the other hand, are of much smaller area (of order 0.01 µm$^2$), and their critical currents are relatively insensitive to applied magnetic fields, varying by no more than 2% in fields of 0 ± 5mT [9].

The critical currents of the nanobridges, however, are temperature dependent, as previously reported [8]. As the operating temperature ranged from 4.5-6.5K in the present experiments, the bias current was adjusted as required to optimise the *voltage amplitude* of the SQUID's V-B characteristic. It was confirmed by repeated measurements at a given temperature, that these adjustments to the bias current did not cause any change in the values obtained for *ΔB*, as defined by (Eq. (2).

In the following Section, we shall describe the use of these single-layer, composite washer-SQUID devices incorporating nano-bridge weak-link junctions to measure the temperature-dependent penetration depth of the inserted washer material

## IV. EXPERIMENTAL RESULTS

The SQUID devices, such as described above, are cooled in a well-shielded enclosure to liquid helium temperatures on a stage which can be temperature-controlled in the range 4.2K-20K with a stability of 0.005K. A 50-turn field coil produces magnetic fields up to 50 µT at the SQUID. Four-terminal dc connections enable conventional *I-V* characteristics to be recorded using a room-temperature amplifier with a gain of 1000, and a white noise floor of ~3 nV/Hz$^{1/2}$. For more sensitive measurements, an ac bias current at 40 kHz is applied to the SQUID and a low-temperature tuned transformer provides optimum coupling to the amplifier. The noise performance of the system is illustrated in Fig. 2(a), which records the SQUID signal due to an alternating field which produces one flux quantum. The response of the SQUID to dc fields is also shown (Fig. 2(b)), where it can be inferred that the effective area increases by about a factor 2 between *T*~4.5K and 6.15K For comparison, identical SQUIDs that were made without the internal washer showed no temperature-dependent change in effective area over the same temperature range.

The data from a series of measurements of this type are plotted in Fig. 3. From these measurements it is clear that a superconducting-normal transition of the internal washer structure takes place at $T_{ca}$ ~ 6.0K. The suppression of $T_c$ in ultra-thin Nb films is a well-established effect and possible mechanisms are discussed in Ref [5]. In separate resistance-temperature measurements we determined that $T_{cs}$ was between 7.8K and 9.5K, depending on the Nb thickness which ranged from 20 nm to 200 nm.

From these results, experimental values of the effective area of the ultra-thin-film washer structure can be derived, and are shown in Fig. 4 (triangle points), in relation to the theoretical curve, based on Eqs(1), (3). The value of $\lambda_0$ which gives optimum fit is 103 nm.. This is greater than the accepted value (70nm) for pure bulk Nb [3], but is consistent with values obtained elsewhere for ultra-thin or impurity-doped Nb films [3,4]. For comparison with the calculations [1], we observe that our values for the normalised penetration depth $\Lambda/b$ range from 0.05 to 0.75 in the normalised temperature range 0.75 to 0.99.

In Table I, we compare our experimentally-derived values of the normalised effective area Eq.(2) with data taken from Brandt and Clem [1], Fig. 14. It has to be noted however, that their calculations were based on a washer of uniform thickness, whereas our structure for practical reasons as discussed above, was contained within a thick, fully-superconducting ring. The Pearl depth $\Lambda$ can also be derived from the data, as shown in the last column of the Table, and its dependence on temperature is shown in Fig. 5.

It is also instructive to compare our data, achieved for essentially zero frequency fields, with the results of Gubin et al.[5] obtained at 23.45 GHz. For example, by interpolating between their Fig.3 results for 8-nm and 20-nm Nb films at $T/T_c$~0.75, we estimate 1 µm for their effective penetration depth, compared with the value $\Lambda$ =1.05 µm (Table 1) obtained for our 15 nm film.

## V. ESTIMATION OF EXPERIMENTAL UNCERTAINTY

On the assumption that Eq. (2) is accepted as an exact definition of the effective area $A_{eff}$ of the SQUID, the experimental uncertainty associated with measurements of $A_{eff}$ is primarily determined by the precision with which the increment of magnetic field $\Delta B$ corresponding to a change in the SQUID's voltage-field characteristic by exactly one period (e.g. in Fig 2), can be measured. To carry out an evaluation of the experimental uncertainty in the measurement of $\Delta B$ at each of a series of temperature settings, the applied magnetic field was smoothly increased such that the SQUID's sinusoidal voltage-field characteristic underwent an exact number of periods, e.g. 10. From repeated measurements, the mean value of field increment required for exactly 1 period, $\Delta B$, was calculated with an estimated uncertainty of ±3%. The temperature stability during each set of measurements (±5mK) was probably the main cause of random error, especially at temperatures close to $T_{ca}$. From these observations, the random experimental uncertainties in the measurements of $\Delta B(T)$ (Fig. 3) and hence of $A_{eff}(T)$ (Fig. 4), are estimated to be ±3%. Because the values of penetration length $\Lambda(T)$, derived from $A_{eff}(T)$ by means of Eq. (4), are subject to errors in the dimensions *a* and *b*, a systematic error of ~ ±10% is attributed to the *absolute* values of $\Lambda(T)$, and also, therefore, to the figure of $\lambda_0$ = 103nm derived by curve-fitting to the points in Fig. 5. The *changes* in $\Lambda(T)$ with temperature, however, are unaffected by errors in *a* and *b*, so a random uncertainty of ±3% is assigned to the *relative* values of $\Lambda(T)$ in Fig. 5.

## VI. DISCUSSION

As an illustration of a specific case where experimental data on the effective area and/or penetration lengths might be useful in the design and fabrication of a device, we consider the Inductive Superconducting Transition Edge Detector, recently proposed for applications in single-particle and photon detection [10] which depends for its operation on the magnetic and thermal properties of ultra-thin, small-area superconducting structures. The device responds to the modulation of penetration depth $\Lambda(T)$ in an ultra-thin superconducting absorber, under the influence of heating by incident photons or particles. The absorber is in the form of a superconducting patch, located within and inductively coupled to a surrounding SQUID ring. In the presence of an applied perpendicular magnetic field, any change in temperature causes an equivalent change in the effective area of the absorber, and a corresponding change in the SQUID's output voltage. Advantages are gained in terms of an increased energy sensitivity of the SQUID, and a reduced heat capacity of the absorber, by minimising the dimensions of the structures. Moreover, calculations show [10] that the energy sensitivity is optimised by operating the device in the regime where the penetration length is much greater than the absorber dimensions, and where $d\Lambda/dT$ is a strong function of temperature. Based on the measurements described in Section IV above (e.g. comparing curves (i) and (ii) in Fig. 4), it is clear that $d\Lambda/dT$ is strongly dependent on the radius of the inner hole, and so, by reducing it from, say, 5 µm to 2 µm a significant improvement in the overall sensitivity of the device could be achieved. The principle of detection requires that an arbitrary half-integer number of flux quanta is applied to the SQUID, and it would appear that large gains in sensitivity can in principle be achieved by the application of arbitrarily-large magnetic fields. To achieve such an enhancement in practice, however, a much more detailed study of noise, flux creep, and stability will undoubtedly be required, utilising the techniques described in the preceding Sections.

## VII. CONCLUSION

A method has been demonstrated for the experimental measurement of the effective area of an ultra-thin superconducting film structure in terms of the magnetic field increment $\Delta B$ required to introduce one flux quantum $\Phi_0$ into a SQUID ring which surrounds the structure. The particular geometry chosen for the demonstration is that of a slit washer, in recognition of its wide use in the design of SQUIDs, but the principle is more generally applicable in situations where the London penetration depth $\lambda$ greatly exceeds the film thickness, and the Pearl length $\Lambda$ exceeds one or more lateral dimensions of the structure. The method has broad-band sensitivity to magnetic fields from dc up to frequencies of tens of MHz, and is therefore complementary to other well-established techniques. In the reported experiments the effective area was shown to increase by a factor of 2 over the temperature range $0.7<T/T_{ca}<0.95$. Data derived from these experiments for the penetration length of magnetic flux $\Lambda(T)$ and its temperature dependence are shown to be in agreement both with the theoretical calculations of Brandt et al.[1], and with the results of high-frequency measurements of Gubin et al.[5]. As device dimensions become ever smaller, there is a growing need for a straightforward method for the experimental verification of theoretical models. Experimental uncertainties in the measured values of effective area $A_{eff}$, estimated at ±3% were mainly attributed to temperature fluctuations of order ±5 mK. Similar uncertainties were estimated in the relative values of penetration lengths $\Lambda(T)$. Greater accuracy could no doubt be achieved by improvements in the temperature control of the sample stage and enclosure. The absolute values derived for the penetration lengths were subject to a systematic uncertainty of order ±10%, due to probable errors in the measurement of photo-lithographically patterned thin-film structures. Noise in the SQUID, which is a planar device incorporating nano-bridges fabricated by electron-beam lithography or focussed-ion-beam milling, did not contribute significantly to the experimental uncertainty. This i*n-situ*, non-invasive method which, in its present form is readily applicable to structures a few nanometers thick, and a few square micrometers in area, is seen as a useful tool both for the verification of theoretical models, and in the design and optimisation of practical devices.

ACKNOWLEDGEMENT

We are grateful to C.M. Pegrum, D. Hutson and D. Cox for valuable scientific input. This work was funded by the UK Department of Trade and Industry under the NMS Quantum Metrology Program.

References

[1]     E.H. Brandt, J.R. Clem, Phys Rev B69, 184509 pp. 1-12, 2004.

[2]     A.T. Fiory *et al*., Appl. Phys. Lett. Vol. 52, 2165-8, 1988.

[3]     J.H. Claassen *et al*., Phys.Rev.B44, p.9605, 1991.

[4]     R.F. Wang *et al*., Appl. Phys. Lett. vol.75, pp. 3865-7, 1999.

[5]     A.I. Gubin *et a*l., Phys. Rev. B vol. 72, 064503, 2005.

[6]     M.B. Ketchen, IEEE Trans Mag vol. 27, pp. 2916-9, 1991.

[7]     D. Cox, Nanoelectronics Centre, University of Surrey,GU2 7XH, UK.

[8]     S.K.H. Lam and D.L. Tilbrook, Appl. Phys. Lett. vol. 82, pp. 1078-80, 2003

[9]     S.K.H. Lam, unpublished results.

[10]   L. Hao *et al*., Supercond. Sci. and Technol. vol.16, 1479-82, 2003;
         IEEE Trans. Appl. Supercond. vol. 15, 514-517, 2005.

TABLE I

Comparison of experimental data with theoretical SQUID effective area [Ref.1, Fig. 14], for the case of circular geometry, assuming $\lambda_0$ = 103nm for Nb film thickness 15 nm, inner radius $a$=5 μm and outer radius $b$=20 μm; linewidth $s$=5 um, slit 2.5 μm wide x 20 μm long.

| $t=T/T_c$ | $\lambda = \lambda_0/(1-t^4)^{0.5}$ nm | $\Lambda = \lambda^2/d$ μm | $\Lambda/b$ | $A_{eff}/\pi ab$ (Brandt) | $A_{eff}$ (expt[b]) μm$^2$ | $A_{eff}/\pi ab$ (expt) | $\Lambda$ (expt) (μm) |
|---|---|---|---|---|---|---|---|
| 0.75 | 120.9 | 0.974 | 0.049 | 1.11 | 430 | 1.095 | 1.05 |
| 0.95 | 232.2 | 3.594 | 0.180 | 1.28 | 590 | 1.375 | 3.00 |
| 0.99 | 503.8 | 15.0[a] | 0.75 | 1.41 | 860 | 2.19 | 8.20 |

[a] limited by surrounding ring
[b] corrected for geometrical slit area

Figure Captions

Fig. 1 (a) Sketch of SQUID; black areas represent 20 nm Nb film; grey area, 15 nm Nb film. Dotted line indicates position of weak link junctions.
(b) micrograph of actual device. Overall diameter 50 μm, nano-bridges are ~100 nm wide.

Fig.2(a) Noise spectrum of SQUID output signal with applied magnetic field corresponding to 1 flux quantum rms amplitude at frequency of 3 Hz. The signal:noise ratio is ~90:1.
(b) SQUID voltage vs. *B*-field current: Curve (a), $T$=4.47K. Curve (b), $T$=6.15K. Note the flux period almost doubles over this temperature range.

Fig. 3 Field change *ΔB* required to introduce 1 $\Phi_0$ in the SQUID vs. temperature *T*. The continuous curve is drawn for guidance only.

Fig. 4 Effective area of SQUID vs. *T*.
Triangle points, experimental data; the continuous line is for guidance only.
Dashed lines, calculations based on Eqs. (1) and (3) with (i) $a$=5μm, (ii) $a$=2 μm (see Section VI,Discussion). In all cases, $b$=20μm, d = 15 nm, $\lambda_0$ = 103 nm and $T_c$ = 5.98K.

Fig. 5 Penetration length *Λ(T)*. vs. *T*. The open-circle points are experimental data, the continuous line is for guidance only. Dashed line, calculation based on Eq. (1) with *d* = 15 nm, $a$=5μm, $b$=20μm, $\lambda_0$ = 103 nm and $T_c$ = 5.98K.

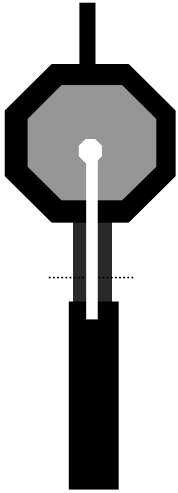 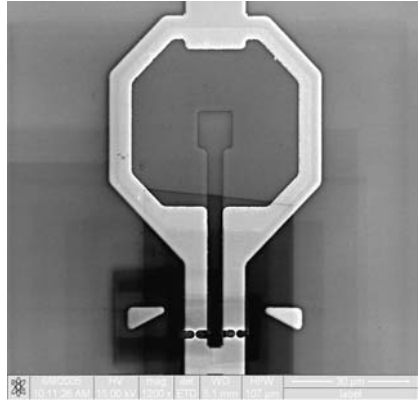

(a) (b)

Fig. 1

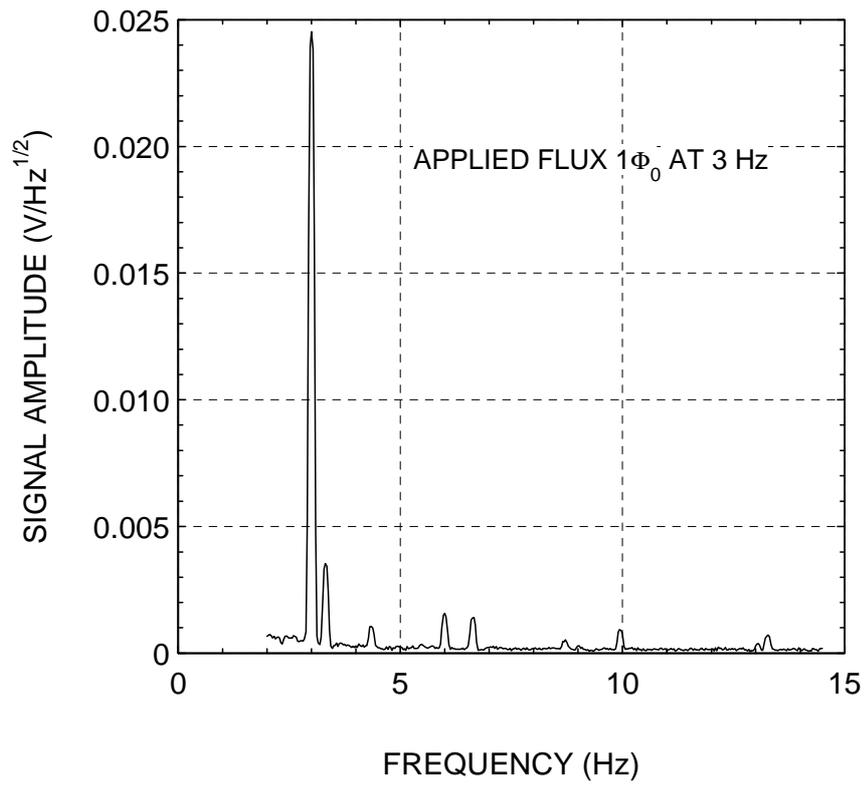

Fig. 2(a)

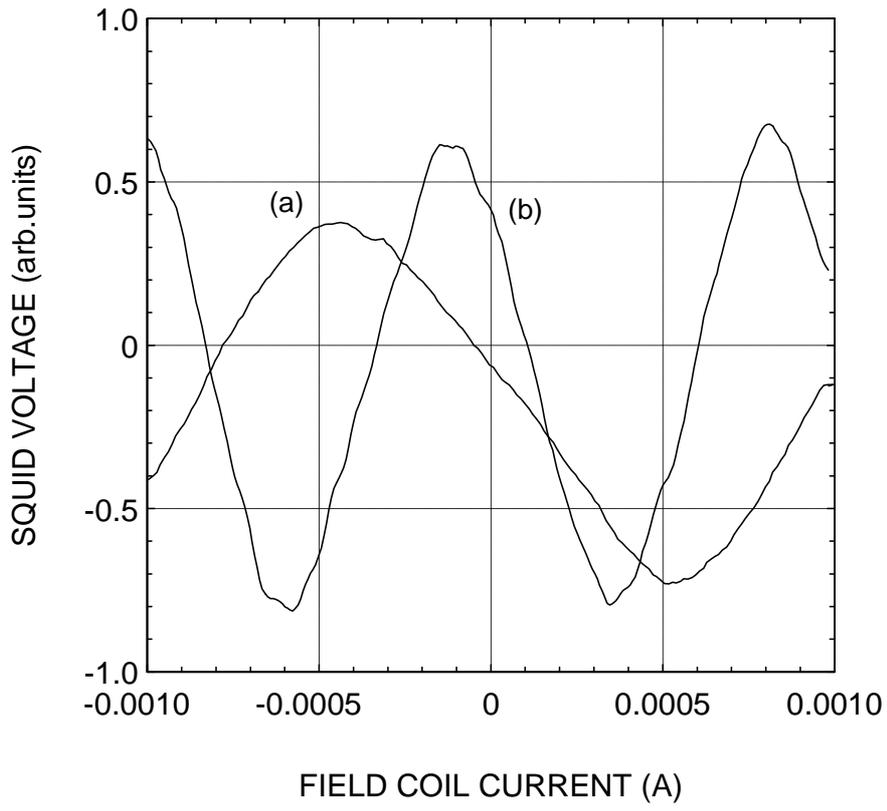

Fig. 2(b)

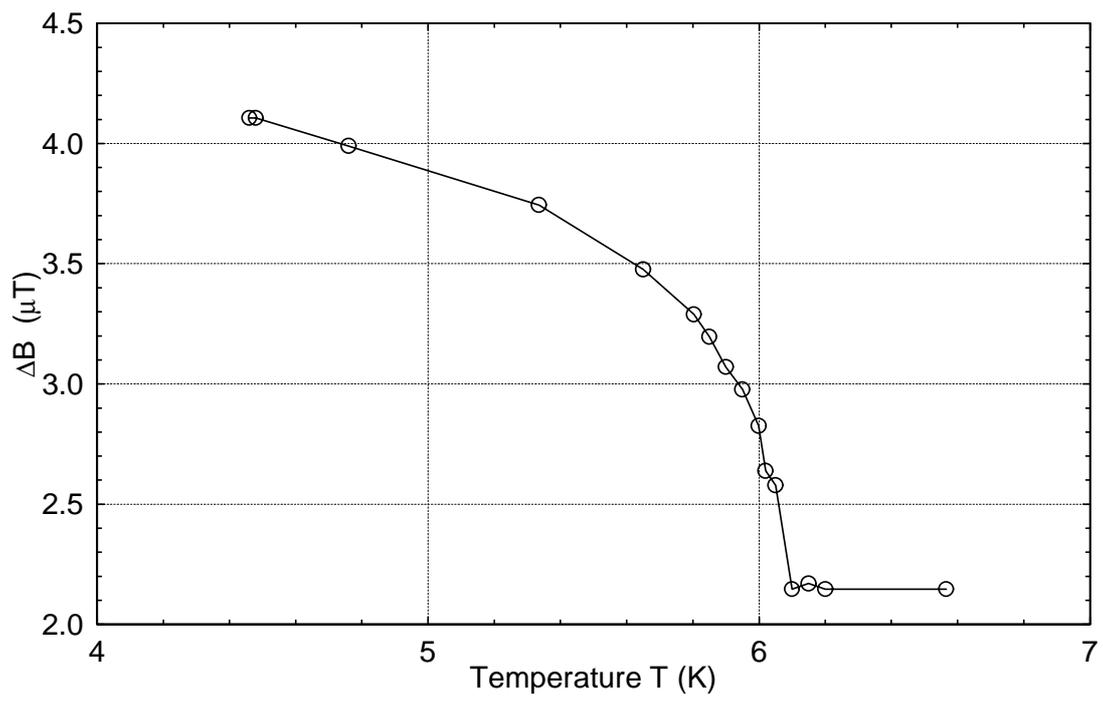

Fig. 3

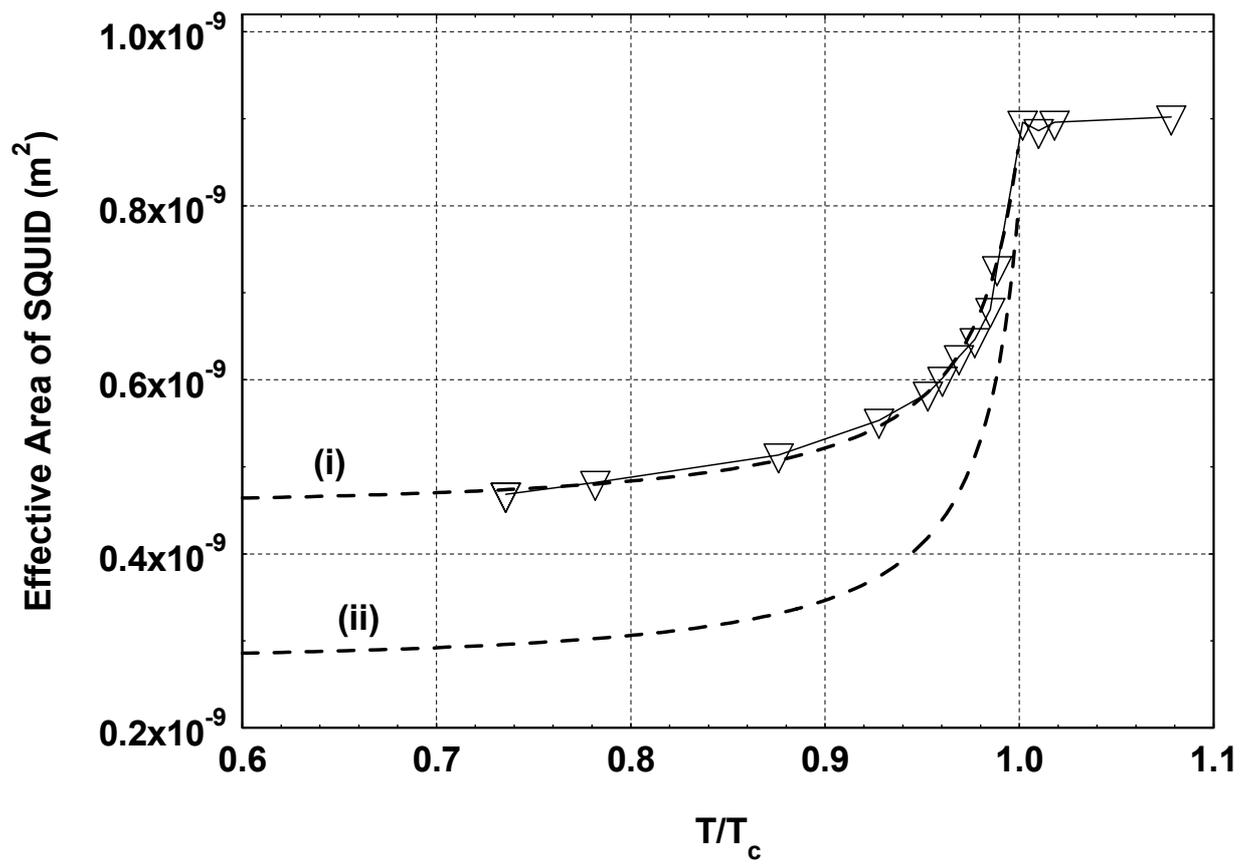

Hao-JAP-Fig. 4 (revised)

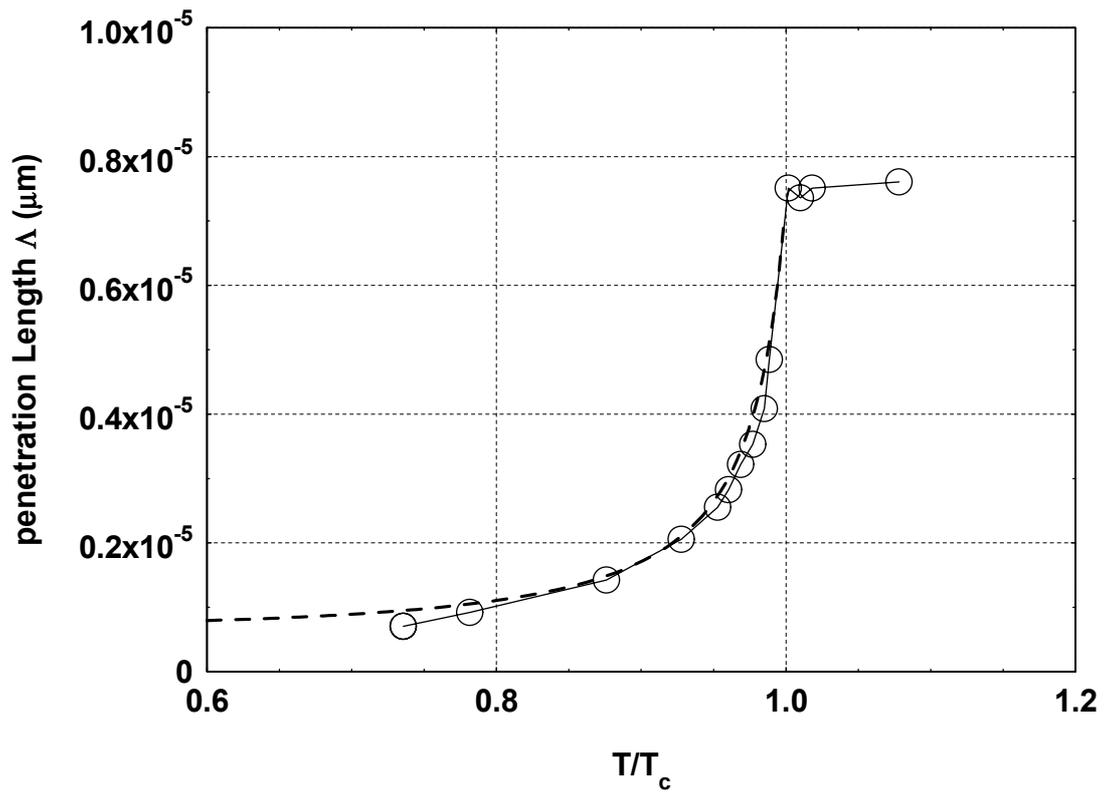

Hao_JAP_revised_Fig5